\documentclass[useAMS,usenatbib,usegraphicx]{mn2e}
\usepackage{amsmath,amssymb}
\usepackage{graphicx,mathptm}
\usepackage{times}
\usepackage{natbib}
\usepackage{psfig}

\newcommand{\Msun}{~\rm{M}_{\odot}}

\newcommand{\pc}{~\rm{pc}}

\newcommand{\Mpc}{~\rm{Mpc}}

\newcommand{\yr}{~\rm{yr}}
\newcommand{\Myr}{~\rm{Myr}}

\newcommand{\MBH}{M_{\rm BH}}
\newcommand{\MBulge}{M_{\rm Bulge}}

\title[Forming massive black holes seeds in galaxy mergers] {Massive black hole
seeds born via direct gas collapse in galaxy mergers: their properties, 
statistics and environment}

\author[Bonoli et al.]
{
Silvia Bonoli$^1$,
Lucio Mayer$^1$ and
Simone Callegari $^{1,2}$
\\ 
$^1$Institute for Theoretical Physics, University of Zurich, Winterthurestrasse 190, 8057 Zurich, Switzerland\\
$^2$Anthropology Institute and Museum, University of Zurich, Winterthurerstrasse 190, 8057 Zurich, Switzerland\\
}

\begin{document}

%\date{}

%\pagerange{\pageref{firstpage}--\pageref{lastpage}} \pubyear{2007}

\maketitle

\label{firstpage}

\begin{abstract}
We study the statistics  and cosmic evolution of massive black hole seeds 
formed during
major mergers of gas-rich late-type galaxies.
Generalizing the results of the hydro-simulations from \citet{mayer10}, we
envision a scenario in which a supermassive star can form at the center of
galaxies that just experienced a major merger owing to a multi-scale
powerful gas inflow, provided that
such galaxies live in haloes with masses above $10^{11} \Msun$,
are gas-rich and disc-dominated, and do not already host a massive black 
hole.
We assume that the ultimate collapse of the  supermassive star leads to 
the
rapid formation of a black hole of $10^5 \Msun$ following a quasi-star 
stage.
Using a model
for galaxy formation applied to the outputs of the Millennium Simulation,
we show that the conditions required for this massive black hole formation
route to take place in the concordance $\Lambda$CDM model are actually common at
high redshift, and can be realized even at low redshift.
Most major mergers above $z \sim 4$ in haloes with mass $> 10^{11} \Msun$ can lead to the formation of a massive seed and,
at $z\sim 2$, the fraction of favourable mergers decreases to about 
half.
Interestingly, we find that even in the local
universe a fraction ($\sim 20 \%$) of major mergers in massive haloes still 
satisfy
the conditions for our massive
black hole formation route. Those late events take place in  galaxies with
a markedly low clustering amplitude, that have lived
in isolation for most of their life, and that are experiencing a major 
merger
for the first time.
We predict that massive black hole seeds from galaxy mergers can dominate
the massive end of the
mass function at high ($z > 4$) and intermediate ($z\sim 2$) redshifts
relative to lighter seeds formed at higher redshift, for example, by the
collapse of Pop III stars.
Finally, a fraction of these massive seeds could lie, soon after
formation, above the $\MBH-\MBulge$ relation.
\end{abstract}


\section{Introduction}

Accreting black holes of masses $> 10^8 \Msun$  are the only known
astrophysical objects able to 
 produce the
enormous amount of energy released by active galactic nuclei
\citep[AGN,][]{lyndenbell69}, and the brightest of these objects, {\it quasars}, are seen at
 redshifts as high as $\sim 7$
\citep{mortlock11}.
The existence of supermassive black holes (black holes with mass $> 10^6 \Msun$) at the center of nearby non-active
 galaxies has been also confirmed 
 in the last few decades thanks to 
 the observation of the gravitational influence of these objects on
the 
surrounding gas and stars \citep[e.g.,][]{kormendy04}.
 
While the existence of black holes in the
cores of most massive galaxies is now well established,  their origin 
  is still largely unknown.   
Several channels of black hole formation have been envisioned
and explored, but when and where the progenitors (``seeds'') of these massive black
holes formed,
is still a topic of intense theoretical
investigation. According to the ``PopIII'' scenario, black holes are the descendant of
 Population III stars, the first generation of stars formed from pristine gas in
dark matter haloes of $\sim
10^6 \Msun$  \citep[e.g.,][]{bond84, madau01, haiman01};
these remnant black holes would have masses of few tens or few hundreds of solar masses,
although recent numerical simulations have shown that the protostellar disk of PopIII stars 
could fragment, lowering the initial mass function of the stars and,
consequently, of their remnant black holes \citep{clark11, greif11}.   
 To grow in a short
time to one billion solar masses (the mass estimated for the black holes powering the
quasars seen
at $z \sim 6$), the Pop III remnants would have to have formed at $z \gtrsim 20$
and accrete gas almost continuously at a rate 
close to the Eddington limit \citep[e.g.][]{volonteri06, tanaka09, tanaka12}.  
Black holes could also come from the runaway collapse of
nuclear star clusters, but even in this case the black hole starting mass would not be
higher than $\sim 10^3 \Msun$  \citep[e.g.,][]{rasio04, portegieszwart02, devecchi09}.

 To relax the tight constraints on growth rates and formation times required to
 build up the very massive black holes powering high-z quasars, the
``direct collapse'' scenario has become more and more attractive, as
 the starting
black hole mass could be few orders of magnitude higher ($10^4 - 10^6 {\rm
M}_{\odot}$). Such massive black hole seeds originate from dense  clouds of gas
 \citep[][]{rees84} at the center of galaxies (or protogalaxies) and likely collapse into a ``supermassive''
star; the supermassive star would then either entirely collapse into a black hole if gravitationally
unstable \citep[e.g.,][]{hoyle63,baumgarte99,shibata02,montero12}, or form a
massive black hole via a ``quasi-star''  \citep[][]{begelman08, begelman10}. 
In the direct collapse scenario, the
most difficult step is to get a massive and dense enough central gas cloud, as
 the gas has to lose its initial angular momentum and reach the
 galactic center before cooling
and fragmentation are able to trigger star formation.  Most works in the
literature have focused on the formation of
black holes from direct collapse in metal-free protogalaxies
\citep[e.g.,][]{lodato06, wise08, regan09, johnson11}, as even 
small traces of metals shorten the gas cooling time significantly  \citep{omukai08}. Even in a
metal-free environment, though, molecular hydrogen cooling has also to be prevented,
which may be possible thanks to fluctuations in the Lyman-Werner background 
\citep[][]{dijkstra08,agarwal12}. 
One could relax the assumption of a metal-free environment if the gas inflow
rate to the center is higher than star formation rate, that is, if enough gas can be
brought
to the center before the bulk of star formation takes place \citep{shlosman89,
begelman09}.

\citet[][hereafter, M10]{mayer10} have recently shown that very efficient central
inflow rates are 
possible in gas-rich galaxy major mergers. Using a set of numerical simulations
with extremely high spatial resolution ($0.1 \rm{pc}$),  M10 found that mergers
can produce  a gravito-turbulent disk in the nuclear
region of the remnant galaxy. The disc is stable against fragmentation, but features
 a strong spiral pattern which supports efficient inflow of gas towards the
galactic center. This strong inflow produces a central rotating cloud of $\sim 10^8 \Msun$
and of the size of few
parsecs, which soon becomes Jeans unstable and collapses to sub-parsec scales. 
The inflow rates from the nuclear disc are so strong (up to few thousands of
solar-masses per year), that this happens in only about $10^5 {\rm yr}$.
The resolution of the simulation does not allow to
investigate further the fate of the nuclear cloud, but the cloud is seen to be
Jeans unstable down
to the smallest resolved scales. This collapsing cloud is thus a likely precursor of
a 
 supermassive star.
Soon after formation, the core of rotating supermassive stars 
 is likely to collapse into a black hole of $\sim 100 \Msun$, as the timescale of nuclear burning is
very short ($\sim 1
\Myr$). In the ``quasi-star'' picture \citep{begelman08, begelman10}, as the
  central black hole starts accreting, the
surrounding gas is inflated into a pressure-supported envelope. While the
envelope loses mass through winds, the black hole keeps accreting at
super-Eddington rates, as the Eddington limit is imposed on the much larger mass of the
envelope, and not on the mass of the black hole itself. \citet{dotan11} estimated that
black holes in this configuration can quickly grow to $10^4-10^5 \Msun$,
provided that their surrounding envelopes  are above $\sim 10^7 \Msun$, as they
would be massive enough for their evaporation timescale to be longer than the
accretion timescale of the black hole. 

Inspired by the numerical results of M10 and the idea of massive black hole
seeds forming from the collapse of nuclear clouds via a supermassive star, in the present work we construct an
analytical model for the formation of massive
seeds in galaxy mergers to be incorporated in a galaxy formation model applied
to the  outputs of the Millennium Simulation. 
Our aim is to study whether the conditions for the formation of a massive seed
  can actually be met in our Universe, what would be the contribution of black
holes from massive seeds to the total black hole population and whether the
descendants of massive seeds could be observationally recognized. 
\citet{volonteri10} made a first attempt to study the cosmological evolution of
massive seeds from
quasi-stars, linking their formation to the mergers of gas-rich 
dark matter haloes and the spin of the haloes, using observational properties of the black hole
population to bracket the values of their model parameters. 
 In the present work we instead directly use 
the results of the M10 hydro-simulations to construct a model for the formation
of massive seeds with essentially no free parameters, whose predictions can be directly compared with data.

In section \S \ref{sec:model} we describe in further details the simulations of M10,
and how we generalize their results to create the analytic model used in the galaxy
formation simulations. Section \S \ref{sec:results} contains the results of
the paper, which we summarize
and discuss in section \S \ref{sec:conclusions}.

\section{Model} \label{sec:model}

In this section we first describe the general properties of the galaxy formation
 formalism upon which we construct the model for the formation and evolution of
black hole seeds (\S \ref{sec:sam}).   
We then discuss the details of our modeling, the origin of the population of
``light'' black hole seeds (\S \ref{sec:light_seeds}) and the details of the 
 hydro-simulations of M10 that inspired our model for the formation of the
``massive'' seed population as described at the end of \S \ref{sec:massive_seeds}.

\subsection{The model of galaxy formation} 
\label{sec:sam}

In the last couple of decades, semi-analytical models of galaxy formation  have
been extensively exploited to study the basic physical processes that drive galaxy
 evolution by comparing the global properties of simulated galaxies with 
observational data \citep[e.g.,][]{kauffmann93,somerville99, bower06}.
Indeed, semi-analytical models offer a simple approach to study large samples of
galaxies, whose formation and evolution is described by a set of coupled differential
 equations  that
combine baryonic physics with the properties of dark matter halo assembly
histories constructed
from the extended Press
\& Schechter formalism or from N-body cosmological simulations. 

The model used in this work uses the dark matter merger trees extracted from
the outputs of the
Millennium Simulation \citep{springel05b}, an N-body simulation
which follows the evolution of cosmic structures in a $500^3 h^{-3} \rm{Mpc}^3$
volume using  $2160^{3} \simeq 10^{10}$ dark matter
particles of mass $\sim 8.6 \times 10^{8} h^{-1} \rm{M}_{\odot}$. The initial
conditions of the Millennium Simulation are based on the cosmological parameters
  of the WMAP1 \& 2dFGRS `concordance'
$\Lambda$CDM framework\footnote{\citet{guo12} showed that there are little differences in
the galaxy properties when a WMAP7 cosmology is assumed}, with $\Omega_{m} =
0.25$, $\Omega_{\Lambda} = 0.75$,
$\sigma_{8} =0.9$, Hubble parameter $h=H_{0}/100 ~ \rm{km} \rm{s}^{-1}
\rm{Mpc}^{-1} =0.73$ and primordial spectral index $n=1$ \citep{spergel03}.
DM haloes and the embedded subhaloes are identified from the output of the
simulation with a
friends-of-friends (FOF) group-finder and an extended version of the {\small
  SUBFIND} algorithm \citep{springel01}, respectively. Dark matter haloes are
considered to be resolved when their mass is above $\sim 10^{10}
h^{-1} \rm{M}_{\odot}$, equivalent to $20$ simulation particles.
The galaxy formation model follows the merger history of dark matter haloes and,
 when a new halo rises above the resolution limit, it gets populated by baryons
in the form of hot gas (according to the cosmic baryon fraction), which will
start  cooling and forming stars
according to the analytical prescriptions adopted. We refer the reader to 
\citet{croton06a} and \citet{delucia07} for all the details on the prescriptions 
for gas cooling, star formation, supernovae feedback and the
other physical processes that shape galaxies and their morphology across cosmic
times, and for a description of the model success in reproducing many
 properties of the observed galaxy population. Here, we modify only the assumptions 
for the formation and evolution of supermassive black holes.

 We consider two different black hole seed formation scenarios. The first one
assumes that seeds originate from relatively {\bf light seeds}, either from massive
Population III stars or from the runaway collapse of
a nuclear star cluster.
This scenario has been already adopted in a number of published works on the
co-evolution of galaxies and black holes and on the clustering of QSOs that are 
based on the same  model of galaxy formation used in this paper
\citep{marulli08, bonoli09}. 
The second channel of black hole seed formation that we introduce in this work is based
on the results of M10 presented above, and aims at describing the formation of
{\bf massive seeds} in major mergers of gas-rich galaxies.

\subsection{Light black hole seeds}  \label{sec:light_seeds}

\begin{figure}
\begin{center}
        \includegraphics[width=0.48\textwidth]{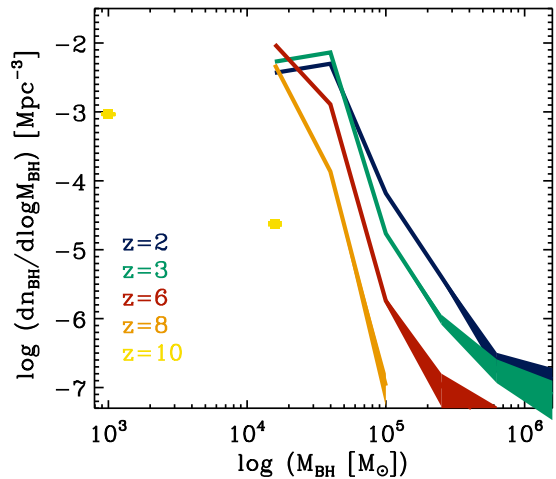}
        \caption{Mass function of the initial black holes, descending from light seeds, that populate newly
formed galaxies. Each curve corresponds to a different redshift of formation of the
host galaxy, as indicated by the legend. Note that at $z=10$ the initial black holes can only take the values $10^3 \Msun$ and
$10^4 \Msun$.}
        \label{fig:Seeds_MF}
\end{center}
\end{figure}

We assume that every newly-resolved galaxy in the simulation contains a black
hole descendant of black hole seeds from PopIII stars or the runaway collapse of
nuclear star clusters, which likely formed in protogalaxies at $z >10$
\citep[e.g.,][]{madau01, devecchi10}. 
Given the limited mass resolution of the Millennium Simulation, we can not
directly track
the formation and early evolution of these objects in our galaxy formation
model, so we have to make some assumptions on the mass of these ``light'' seeds descendants.  
\citet{marulli08} and \citet{bonoli09} 
had assumed a fixed mass for these black holes
 as the global
properties of black holes and quasars at moderate and low redshift 
 ($\sim z < 5$), as simulated by our galaxy formation model, are essentially
insensitive to the mass  assigned to black holes in newly-formed galaxies, since
any subsequent
exponential growth washes out traces of the starting mass. However, as we will
discuss in detail below, our new model for the formation of massive seeds in
galaxy merger requires an accurate census of the mass of black holes, descendant
of light seeds, at any time. 
For this work, we then decided to assume that the mass of the descendant of {\it light seeds} in newly formed
galaxies 
 depends on the mass of the newly-resolved halo they reside in and on the
redshift at which the halo is resolved. 
We use the results of a higher resolution simulation, the Millennium-II\footnote{The
MillenniumII is an N-body simulation with the same
particle number and cosmology of the Millennium, but $125$ times better mass
resolution since run in a volume of $100^3 h^{-3} \rm{Mpc}^3$.}, to construct a
``library'' of typical black hole masses, as a function of halo mass and redshift. 
Thanks to the much higher mass resolution of the Millennium-II, mergers of galaxies residing in
subhaloes down to few times  $10^{8} h^{-1} \rm{M}_{\odot}$ are properly followed. 
\citet{guo11}
have generated a galaxy catalogue for this simulation, with black holes forming
and growing during galaxy mergers as in \citet{croton06a}, and we use this
catalogue\footnote{from
the online database {\it http://www.mpa-garching.mpg.de/Millennium/} \citep{lemson06}} to generate our
``library''. To compensate for still limited resolution of the Millennium-II, 
we impose a minimum mass of $10^3 \Msun$ for
these light black hole seed descendants (but we note that the results presented in this paper are insensitive to 
the exact value of the minimum black hole mass imposed).
In Figure \ref{fig:Seeds_MF} we show, at various redshifts, the mass function of
the black holes populating newly resolved galaxies, which are assumed to be the
descendants of light seeds. At $z=10$ the typical assigned masses are $10^3$ or
$10^4 \Msun$. As redshift decreases, the mass assigned to new black holes is
higher 
and, for galaxies resolved at $z <3$, it can even be higher than  $10^6 \Msun$. 

These black holes will keep growing mainly during
 galaxy mergers.  
When a merger takes place, we assume that
the black holes hosted by the two galaxies coalesce instantaneously (so that each galaxy hosts only one
black hole at a time), and the resulting black hole starts
accreting a fraction $\Delta M_{\rm BH}$ of the surrounding cold gas  $m_{\rm cold}$
which depends on the mass ratio of
the merging galaxies ($m_{\rm sat}/m_{\rm central}$)  and the redshift of the
merger  \citep[as in][]{croton06b}. Specifically,
the amount of gas accreted by a black hole during a galaxy merger is given by:
\begin{equation} \label{eqn:quasar_merg_z}
\Delta M_{BH, \rm merger} = \frac{f'_{\rm BH} \ m_{\rm cold}}{1 +
  (280\,\rm{km\,s^{-1}}/V_{\rm vir})^2} (1+z_{\rm merg})  \, , 
\end{equation}
where $m_{\rm cold}$ is the total mass of cold gas in the final galaxy and 
$ f'_{\rm BH} = f_{\rm BH}\ (m_{\rm sat}/m_{\rm central}) $ and $f_{\rm
BH}\approx 0.02$  is a normalization parameter chosen to match the observed local
$\MBH-\MBulge$. This prescription has been shown to be successful in
reproducing not only the $\MBH-\MBulge$ relations, but also the black hole mass
function at $z=0$ and other observed properties of black holes, quasars and
their environment \citep{marulli08, bonoli09}, and the choice of populating
newly-formed galaxies with the typical black holes found in the Millennium-II does not
modify these results.

\subsection{Massive Black Hole Seeds} \label{sec:massive_seeds}

The model we construct for the formation of massive black hole seeds from galaxy
mergers is  based on the assumptions and results of the hydro-simulations of
M10. We thus first describe their results and then how we generalize them to build a model to be used in the galaxy formation framework described above.

\subsubsection{The evolution of the nuclear region of merger remnants according to
hydro-simulations}

M10 have used a set of numerical simulations to
study the evolution of the nuclear region of high-redshift major-merger
remnants.  The initial conditions of such mergers will be used in the analytical
model that we develop in this paper. The model galaxies adopted in M10 
are
disk dominated galaxies (the bulge-to-disk mass ratio  is typical of 
present-day
late type spirals, i.e. $B/D=0.2$), with a gas/stars fraction in the disk 
of $20\%$
just before the merger occurs (some gas can be consumed by star formation 
during the tidal
interaction preceding the merger, an effect that will depend on the 
details of the orbital parameters
and internal structure of the merging galaxies).
The choice of the gas fraction is conservative since high-z disks can be 
significantly more gas-rich
 \citep{genzel06}.
Only equal mass mergers of galaxies were considered, with a dark host halo mass range 
between
$5 \times 10^{10}$ to $10^{12} M_{\odot}$. At lower mass scales outflows 
driven by supernovae
feedback should dominate the gas dynamics and thermodynamics, preventing 
central
accumulation of gas \citep{governato10}. Mergers with mass ratio below 
$1:3$ have been
shown to be much less efficient at concentrating gas to the inner few tens 
of parsecs
\citep{callegari09, guedes11}, and were therefore not 
considered in M10 nor they will be in the  model formulated in this paper. 

Thanks to the unprecedented spatial resolution of their experiments (the
softening length of the gas in the nucleus is $0.1 \;\rm{pc}$), MA10 
find that major mergers of disky and gas-rich galaxies in haloes
above $10^{11} \Msun$ are responsible for a very strong 
inflow
of gas down to scales of a few parsecs, peaking at 
$ > 10^4 M_{\odot}/\yr$.
A key point in this result is that the resulting inflow rates are a few 
orders
of magnitude higher than those found in simulations
of nearly isolated unstable self-gravitating protogalactic disks \citep[$ \sim 10^3 - 10^4
M_{\odot}/\yr$  vs. a few tens
of solar masses/year, see e.g.,][]{regan09}.
The reason is that in mergers
gas can be shocked and torqued much more effectively, losing most of its 
angular momentum
over a short timescale.

A much higher predicted inflow rate is also a difference with respect 
to semi-analytical models of seed black hole formation via direct collapse
\citep[e.g.][]{volonteri10}, which consider the conditions at the
virial radius of the host galaxy halo. Indeed, in M10 gas flows inward owing
to continuous loss of angular momentum by  torques and shocks over
a wide range of spatial scales (at small scales the torques are provided by spiral 
arms in the unstable nuclear disk arising at the center of the merger remnant).
The net inflow rate at the center thus depends on the integrated effect of all
the torques at different scales rather than on the conditions at the global
galactic or halo scales. The resulting characteristic velocity of the gas is
much higher than the free fall velocity at the halo virial radius, while the
gas mass reservoir is still within a factor of a few of the total amount of gas
available in the galaxy even at relatively small scales (i.e. the nuclear disk
acquires most of the cold gas mass in the galaxy due to angular momentum loss
at large scales). Ultimately, this translates into a much higher mass inflow rate
that well exceeds the expectation using free fall velocity at the 
scale of the halo. The latter is given by:
\begin{equation}
dM/dt = m_{d} {V_c}^3/G,
\end{equation}
where G is the gravitational constant, $V_c$  is the virial circular
velocity of the dark matter halo hosting the galaxy and $m_d$ indicates the fraction of mass 
which is able to collapse in free fall. For $V_c \sim 150 - 200
\rm{km/s}$, which were the typical values of the virial velocities of the haloes simulated 
 by M10, one would obtain $dM/dt \sim$ few tens of $\Msun/ \yr$  (assuming $m_d
=0.05 - 0.1$ as in \citet{lodato06}), which is several
orders of magnitude lower than what is found in the sub-pc scale merger 
simulations.
Likewise, while \citet{volonteri10} \citep[see also][]{lodato06, volonteri08}
use a threshold in spin parameter of the halo as
a further criterion to decide which haloes would undergo seed formation via
direct collapse, we avoid that based on the fact that the angular momentum
of the gas likely does not trace that of dark matter at any scale \citep[e.g.,][]{dubois12}, 
and especially at small scales the dynamics of the nuclear
disk dominates the evolution of the angular momentum.

Shortly after the merger,  the central region
of the galaxy remnant in M10 is characterized essentially by 
two components:
\begin{itemize}
\item {a rotating nuclear disk, of the size of about $\sim 80 \pc$ and 
mass
of $\sim
10^{9} \Msun$. It is in this disk where a very large fraction of 
the gas
initially in the galactic disk has accumulated;}
\item {a pressure-supported rotating cloud of the size of few parsecs, and 
of
mass of  $\sim 10^{8} \Msun$. This cloud forms from the rapid 
inflow
of gas from the nuclear disk less than a million years after the merger.
  Due to
this strong inflow from the surrounding disk, the cloud becomes Jeans 
unstable,
and eventually collapses to sub-parsec scales.}
\end{itemize}

Once the cloud shrinks to the resolution limit ($0.1 \pc$) the simulation 
cannot
follow the collapse further, hence whether it forms a supermassive star 
first or collapses
all the way into a seed black hole is uncertain at this stage. Both 
processes, however,
would lead to the formation of a massive seed on timescales short enough 
($< 10^7  \yr$  even
in the case of the slower route via a supermassive star stage) that, for 
the purpose of this
work we will consider  instantaneous. The mass of the seed 
in the runaway
gravitational collapse case could be much larger, comparable with the mass 
of the cloud,
but we will consider the supermassive star route in order to be 
conservative ($M \sim 10^5 \Msun$,
see \citet{begelman10} and, for the cases of 
the large inflow
rates considered here, \citet{dotan11}).
The circumnuclear disk + supermassive cloud configuration is the 
underlying structure that we will consider to construct our 
phenomenological
recipe of massive seed formation and growth.

\subsubsection{Massive black hole seeds in the galaxy formation model}
\label{sec:DC_criteria}

In our new model, we  
assume that massive black hole seeds form during the major mergers of massive gas-rich
late-type galaxies that satisfy specific constraints.
 Mimicking the initial conditions used in the simulations of M10, the exact 
conditions for inserting a massive seed are the
following: 
\begin{enumerate}
\item {\bf A major merger.} We impose a minimum mass ratio $M_{gal,1}/M_{gal,2}$ of 0.3,
where with $M_{gal}$ we refer to the stellar$+$cold-gas component of the galaxy.
As discussed above, this is  motivated by simulations results \citep{callegari09, guedes11}
who showed that
mergers of galaxies with smaller mass ratios are not violent enough to trigger
strong gas inflows towards the galaxy center;
\item {\bf A minimum mass of the dark matter halo of the merger remnant
$\mathbf{M_{halo}}$ of $\mathbf{10^
{11} \Msun}$.} $M_{halo}$ corresponds to the subhalo mass for satellite galaxies and the virial mass for centrals . 
 This threshold
 in halo mass comes from the results of 
the M10 simulations, which show,  with varying mass of the merging
galaxies, that the 
central collapse does not occur below such mass scale as a result of 
increased stability of the nuclear disk.   Future analysis of an extended sample of 
numerical simulations will have to clarify if it is indeed the mass or rather
the central density of 
galaxies, as probed by the maximum circular velocity $V_{max}$,  the
fundamental variable governing the mass inflow.
\item {\bf A bulge to total ratio $B/T$ of both merging galaxies of at most
$\mathbf{0.2}$}. The results of M10 clearly show the importance of a disk to sustain the
spiral patterns able to feed the nuclear region of the merger remnant. Galaxies
with such low $B/T$ ratio have both very large gas fractions, so they
potentially have
 enough fuel to feed the nuclear region of the merger remnant, and the structure
for a  proper
dynamical response to generate the required multi-scale gas inflow as a result of
tidal interactions (a dominant bulge tends to stabilize a galactic disk even in the
presence
of strong tidal perturbations, weakening the gas inflow).
\item {\bf The lack of a black hole of mass larger than $\mathbf{10^{6}
 \Msun}$.} As pointed out by M10, any pre-existing massive black hole fed by the inflowing gas  
  would stabilize the nuclear disk via radiative feedback. We set $10^6 \Msun$ as threshold, as the feedback from smaller black holes would likely be too weak to halt the gas inflow and only massive black holes might be able to actually reach the center remnant, thanks to dynamical friction, in timescales shorter than the timescale of gas inflow and formation of the central cloud \citep[][]{mayer07,chapon11}. 
\end{enumerate} 
If the above conditions are met, we assume that the 
merger is able to give rise to the high gas inflow rates that lead to the post-merger configuration found by M10: a 
nuclear cloud of $10^{8} \Msun$ surrounded and fed by a circumnuclear
disk. We assume that the central part of the cloud 
quickly collapses into a supermassive star
which, in a short timescale, gives rise to a black hole of $\sim 10^5 {\rm
M}_{\odot}$ via a ``quasi-star''. As the timescales for these processes are very
short compared to the time resolution of the cosmological simulation, we assume
that a massive seed of $10^{5} \Msun$ is formed as soon as the conditions on the
merger described above
are met. 

After being formed, the seed can start accreting the gas still flowing from the
circumnuclear
disk to the surrounding residual cloud. We assume the circumnuclear disk to be
$2/3$ of the total mass in  cold gas, which is
approximately the value found by M10.
Such high fractions of gas  in the central region of  major merger
remnants have also been seen in previous simulations \citep[e.g.,][]{barnes96}. 
Further assuming that stars in the disc form with a $30\%$ efficiency (which can
be seen as an upper limit on the star formation efficiencies in molecular clouds
at small scales, i.e. well below $100 \pc$, e.g., \citet{evans09}),
about $1/3$ of the gas in the nuclear disc can feed the remnant nuclear cloud and then
be available for the black hole.

\begin{figure}
\begin{center}
        \includegraphics[width=0.48\textwidth]{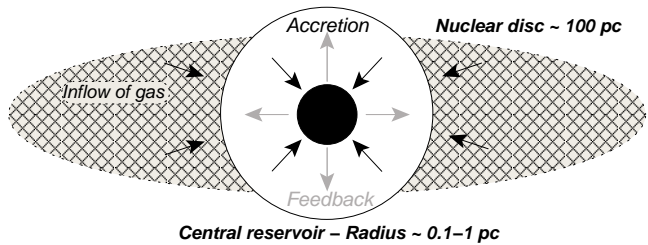}
        \caption{Sketch of the structure surrounding the massive black hole seed
(of $\sim 10^5 \Msun$)
formed from the supermassive star after the quasi-star phase. The reservoir of
gas from which the black hole
grows includes the remnants of the central massive cloud from which the initial
supermassive star formed and the gas still flowing from the nuclear disk. The
growth of the seed stops once its feedback energy balances the binding energy of
the central reservoir. }
        \label{fig:reservoir}
\end{center}
\end{figure}

With these numbers in mind, we assume that the remnant of the nuclear cloud and the gas still
 flowing from the nuclear disk form a ``reservoir'' of gas of  $M_{r}=
2/9 M_{gas}$ (with $M_{gas}$ being the total cold-gas gas in the galaxy), from which the 
newly-formed massive
black hole seed can accrete. The black hole now grows at much lower rates
than the super-Eddington rates that made its fast formation possible during the quasi-star phase, 
and it keeps growing  until its feedback energy is able to
unbind the reservoir (see the sketch in Figure \ref{fig:reservoir}). 

Equating the total energy emitted by the black hole during the
accretion phase and the binding energy of the gas reservoir, we get:
\begin{equation}
\label{eqn:Eres}
\int_{t_i}^{t_f}\epsilon_{feed} \epsilon_{r} \dot{M}  c^2 dt = \frac{G
M_{r}^2}{r_r},
\end{equation}
where 
$\epsilon_{feed}$  is the efficiency of feedback coupling, which we set to
$0.05$ as the value typically adopted by the community \citep[e.g.,][]{diMatteo05}, 
$\epsilon_{r}$  is the radiative
efficiency,  $\dot{M}$ is the accretion rate, and $r_r$ is the radius of the region.
The amount of gas accreted by the  black hole seeds is then given by:
\begin{equation}
\label{eqn:DM_DC}
\Delta M_{BH, reservoir}= \frac{1}{\epsilon_{feed} \epsilon_{r} c^2} \frac{G
M_{r}^2}{r_r}.
\end{equation}
We assume that $\Delta M_{BH, reservoir}$ is accreted at the Eddington rate of
the black hole.
However, given the uncertainties on the properties of the ambient medium, we
do not try here to estimate the luminosity output of those events, and we
postpone this further modeling to a future work. 

The model developed here relies on few parameters, whose values are set by the
 hydro-simulation results of M10. The only parameter not well constrained by the
simulations is $r_r$, the radius of the central reservoir.  We decided to assume two values for this
parameter, $r_r = 1 {\rm pc}$  and $r_r = 0.1 {\rm pc}$, which are plausible
assumptions in the range of the characteristic sizes of the supermassive clouds
found in M10.  
Given the importance of  $r_r$  in setting the final mass of massive
seeds after their first accretion phase, most of the results  in the next sections
will be discussed for both values of this parameter. 

After the formation and having accreted from the reservoir cloud, the massive black
hole seeds can further grow  when their host galaxy  experiences a new merger.
In those subsequent accretion events, the massive black hole seeds are treated in the same manner as the
descendants of light seeds: they merge with the black hole hosted by the companion galaxy and 
  grow according to equation
\ref{eqn:quasar_merg_z}. We use this  prescription for black hole  growth
for all the mergers that do not satisfy the condition for
the creation of a massive seed, as we do not have at the moment enough input
from small-scale hydro-simulation to attempt a general new recipe that would hold
for all kinds of mergers.

Finally,  if a massive seeds merges with
a light seed, the new black hole will have
the ``light'' or ``massive'' seed tag, depending on which black hole progenitor is larger.

\section{Results} 
\label{sec:results}

After having set the theoretical assumptions that define our model, we now look
at how frequently the imposed conditions for the formation of a massive seed are
met as galaxies evolve from high redshift to the local universe (\S \ref{sec:statistics}). We then look at 
 the relative importance of black holes from massive and
light seeds in defining the total black hole population (\S \ref{sec:properties}).
Finally, we  compare the typical environment in which the descendant of light
and massive seed reside, 
trying to find differences which could help tracing back the
 origin of black holes (\S \ref{sec:environment}).

\subsection{Statistics of events} \label{sec:statistics}

We first of all explore if and how frequently the conditions for the
formation of a massive seed from galaxy mergers described in section \S
\ref{sec:DC_criteria} are actually met in a realistic $\Lambda$CDM universe.

\begin{figure}
\begin{center}
        \includegraphics[width=0.48\textwidth]{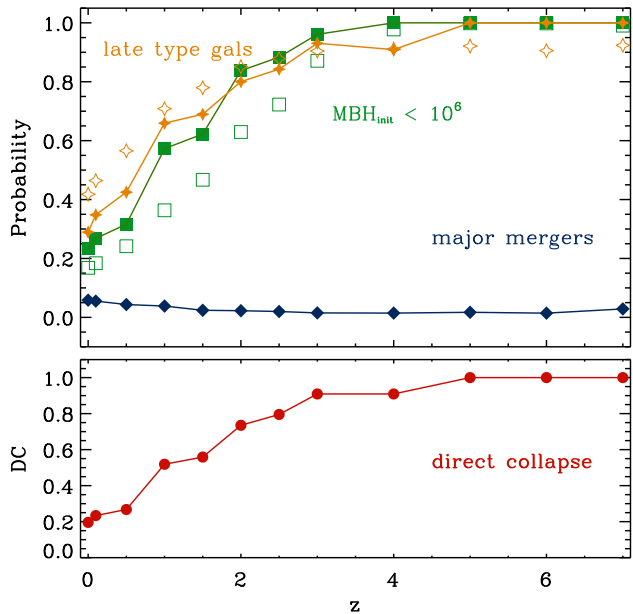}
        \caption{For the merger events of galaxies hosted by haloes of at least
$10^{11} \Msun$, probability of satisfying the conditions imposed for the formation of a massive
seed via direct
collapse: the blue diamonds show the fraction of major mergers, the green 
squares the fraction of merging galaxies with a black hole smaller than $10^5
\Msun$, and the orange stars indicate the fraction of mergers that
involve late-type galaxies. The red bullets in the lower panel indicate the fraction of major
mergers that satisfy all the conditions for the formation of a black hole seed
from direct collapse. The empty symbols indicate the fraction of galaxies with a
small black hole and with late-type morphology for major mergers only.} 
        \label{fig:constraints_z_evol}
\end{center}
\end{figure}
 
\begin{figure}
\begin{center}
        \includegraphics[width=0.48\textwidth]{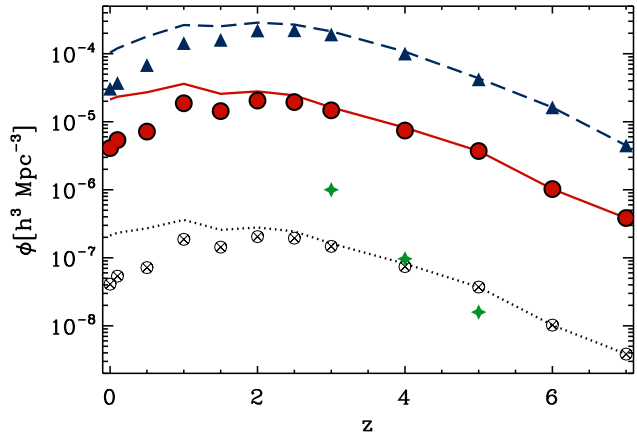}
        \caption{Number density of major mergers (mass ratio $> 1:3$) of
galaxies hosted by subhaloes of
at least $10^{11} \rm{M}_{\odot}$, obtained from the rate of events and assuming a
``visibility time'' of $100 \Myr$
(red solid line). Of these mergers, the number density of the ones that satisfy
the requirements for the
formation of a massive seed from direct collapse is indicated by the large red
bullets.
The blue dotted line and the blue triangles show the same quantities, but
assuming that the threshold for major mergers is $1:10$ instead of $1:3$. 
Finally, the black dashed line and the crossed black circles show again the
number densities of major mergers (with $1:3$ mass ratio threshold) and events of massive seed formation, but
assuming a ``visibility time'' of $1 \Myr$.  The green stars indicate the number density of observed
 high-redshift quasars from \citet{shen07}, put here to provide the reader with an order-of-magnitude
comparison.} 
        \label{fig:Ndensity_MM}
\end{center}
\end{figure}

In the top panel of Figure \ref{fig:constraints_z_evol}, we show, as a function of
redshift, the probability of satisfying each of  our condition for
the formation of a massive black hole seed. Of all mergers of galaxies whose remnants is hosted by a
subhalo of at least  $10^{11} \Msun$,  the fraction of major mergers (mass ratio $1:3$, blue 
diamonds), is approximately
constant with redshift, and of the order of few percents. 
The fraction of merging
galaxies hosting a black hole (descendent of a light seed) smaller than $10^6
\Msun$ is, as expected, a strong function of redshift (green squares): at high
redshift, the probability that a major merger involves galaxies with a small black
hole is quite high (and even equal to one above $\sim z=4$), but it
decreases sharply at lower redshift and, in the local universe, about 
$20\%$ of both merging galaxies have a black hole smaller than $10^6 \Msun$.
If the merging galaxies are experiencing their first merger, the mass of the black holes they host 
is still the mass assigned when the host galaxies were first formed.  In this work, the mass assigned to these black holes, assumed to be the descendants of light seeds,  is estimated using the outputs of the Millennium II simulation, as
discussed in  section \S \ref{sec:light_seeds}. If we had assumed a fixed mass of $10^{3}-10^{4} \Msun$, as was done in previous works, the probability of satisfy the constrain on the pre-existent black hole mass would have been higher, in particular at low-z. Moreover, it might be plausible that
the descendant of light seeds do not populate all galaxies, as we conservatively assume here, as some POP III stars might never reach high enough masses to form a black hole \citep[see, eg., the discussion in ][]{volonteri12}.  In this respect, then, the numbers shown here can then be considered as lower limits.
On the other end, a channel of black hole feeding not considered in the reference model which could make black hole masses rise significantly  before any merger takes place, is growth by secular instabilities. 
  We tried to add this growth channel to look for any change in the statistics of events of massive seed formation found with our reference model. Our reference galaxy formation model already has a prescription\footnote{The stability criterion used in the model is the one introduced by 
 \citet{mo98}: the stellar disk of a galaxy becomes unstable when
this inequality is met: 
$ \frac{V_c}{({\rm G} m_{\rm{disk}} / r_{\rm{disk}})^{1/2}} \le 1$.
If the disc is unstable (its mass in stars is larger than the mass that gives rise to the inequality), 
the excess stellar mass is transferred from the disk to the bulge to restore
stability.} for testing the instability of galactic disk: we simply assumed that when instability is detected, not only the bulge grows, but also a small fraction of the gas in the galaxy reaches the nuclear region and feeds the central black hole. While we find that the additional growth by disk instability  
 has a detectable effect in increasing the  total normalization of the black
hole mass function, it does not  influence 
the statistics of massive seed formation. Disk instability events, in fact, mainly take place is galaxies that never satisfy the other conditions for the formation of a massive seed: as discussed by \citet{guo12}, disk instability is important for building up bulges in Milky Way-type galaxies, which we do not expect to be hosting a massive seed. 

 The other constraint we impose for the formation of a massive seed is the low bulge-to-disc ratio of the merging galaxies. The probability of having two disk
galaxies involved in a merger is shown in the same figure (orange stars). 
The fraction of merging disk galaxies is also quite high (between $0.8$ and $1$)
above $z=3$, but
decreases with decreasing redshift, as the total fraction of disk galaxies is
lower. The morphology in the code is treated
as in \citet{croton06a} and \citet{delucia07}; in a later development of the model
introduced by \citet{guo11}, galaxies suffer a more gradual stripping of cold
gas when they become satellites, with respect to the previous models. In this
case satellite galaxies have even lower B/T ratio at all z, as the gas remains
on the satellite and can keep
cooling and forming stars in the disk, and central
galaxies also have, on average, lower B/T disk ratio. With this newer version of the model, the 
fraction of merging galaxies with low B/T ratio would even be higher, and our estimates on the probability of satisfying the condition on the galaxy morphology can also 
be regarded as lower limits.

The resulting fraction of major mergers that
satisfy all the conditions to form a massive seed black hole is shown in the lower panel of the
 same figure: all major mergers
above $z \sim 4$ could potentially lead
to the formation of a massive black hole from direct collapse. At $z \sim 2$ the
fraction of major mergers that can form a massive seed has decreased to half,
and in the local universe only $20 \%$ of major mergers lead to direct collapse.
Yet, it is particularly interesting that our model predicts the presence of favourable
conditions for the formation of black holes from direct collapse  in a
small fraction of major mergers in the local Universe. While we defer to 
future work a detailed study of the possible observational signatures of these
events, in  section \S \ref{sec:environment}
we briefly discuss some properties of their environment.

\begin{figure*}
\begin{center}
        \includegraphics[width=0.9\textwidth]{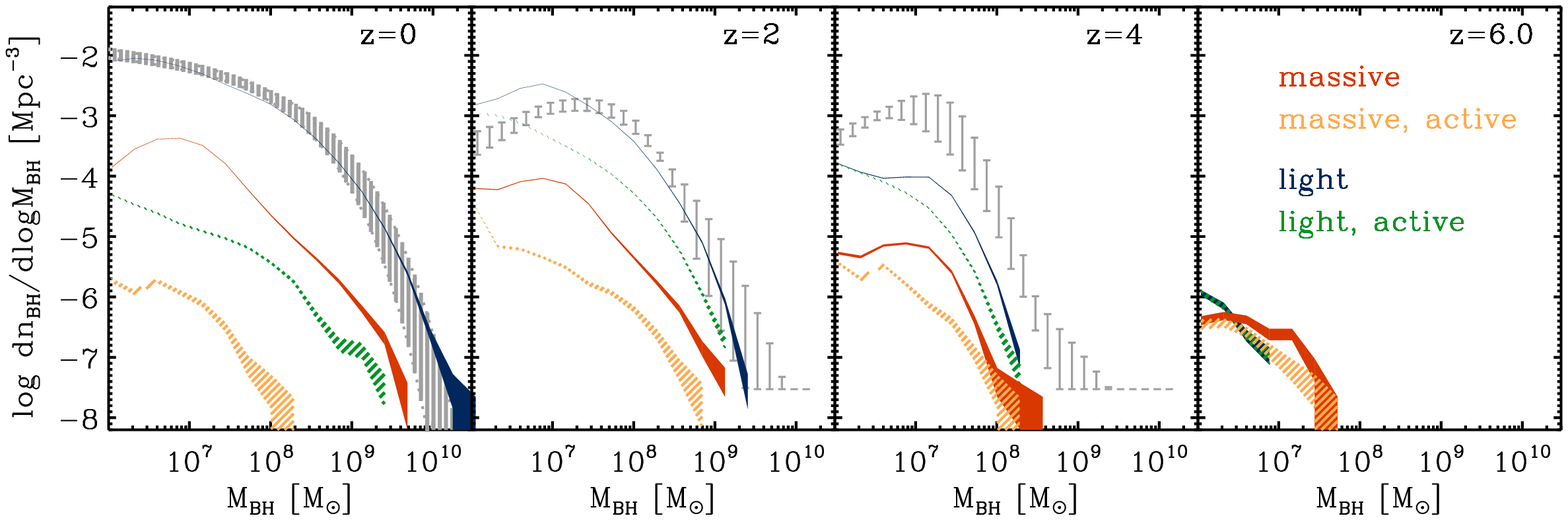}
        \includegraphics[width=0.9\textwidth]{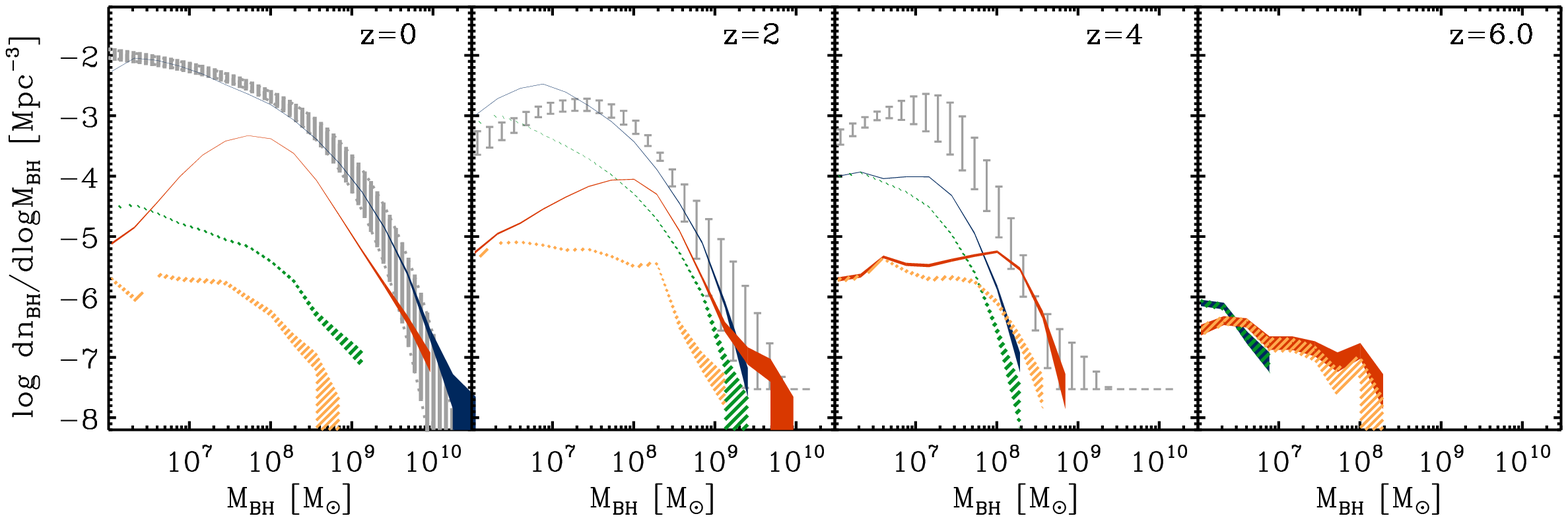}
        \caption{Black Hole mass function at various redshifts as indicated in
the panels. For the top panels we have assumed that newly-formed massive seeds grow from a gas reservoir of  $1  \pc$, 
while for the bottom panels a radius of $0.1  \pc$ is assumed.  The blue curve shows the mass
function of black holes with a light progenitor, while the dark red curve shows
the mass function of black holes with progenitor from direct collapse. The
dashed curves indicate the mass functions of black holes emitting at more that
$10\%$ the Eddington limit, for the light seed descendants and massive seed back
holes
(light green and orange respectively).  The width of the lines include the $1-\sigma$
error.  At $z=0$, the grey band shows the
black hole mass function calculated by \citet{shankar04}, while at $z=2$ and at
$z=4$ the grey band is the estimate for the mass function derived through a
continuity equation by \citet{merloni08}. }
        \label{fig:MF_BH}
\end{center}
\end{figure*}

In Figure \ref{fig:Ndensity_MM}, we show the redshift evolution of the number
density of major merger as well as of 
 massive seed formation events. As from our
simulation we have information on the \textit{merger rate}, that is, the number
of mergers per time interval within two subsequent snapshots, to obtain the number density of 
 possibly visible events, we have to assume a typical timescale for the duration of the events. The numbers shown
in the figure are obtained assuming, for each merger, a
visibility time of $10^8 \rm{yr}$, which is approximately the estimated life-time of bright quasars \citep[e.g.,][]{martini04}. Following this definition, the number
density of major merger events of galaxies hosted by subhaloes above
$10^{11} \Msun$ is indicated by the red solid line. The red bullets show the
number density of these major mergers that also satisfy all the condition for
the formation of a massive seed: as discussed above, essentially
all major mergers at high redshift satisfy the conditions for the formation of a
massive seed, while, at lower redshift, fewer major mergers can lead to a new
massive seed.     
To guide the eye with an order-of-magnitude comparison, we plotted in the same figure the number density of high-redshift observed optical quasar reported by \citet{shen07}. 
While a direct comparison between the observed quasar properties and
our black hole population is beyond the scope of this paper, we see that the number of
events possibly forming direct collapse seeds is large enough to account for
bright optical quasars. 
In the same figure, the blue dashed line and the blue triangles show the number
density of mergers and events of massive seed formation if the threshold for a
major merger is decreased from $1:3$ to $1:10$, as used in  \citet{volonteri10}
in their study  of  the
 evolution of quasi-stars. Clearly, the number of events assuming a smaller
mass-ratio threshold is much higher, approximately an order of magnitude higher
at all redshifts. As discussed in section \S 
\ref{sec:DC_criteria}, we conservatively use a mass threshold of $1:3$ in our
reference model as mergers of galaxies with smaller mass ratios might not be
violent enough to lead to the high gas inflow rates necessary to form a
supermassive cloud. 
Finally,  in Figure \ref{fig:Ndensity_MM} we also show the number density of major
merger and massive seed formation events for our reference model, but  
assuming a much shorter visibility time ($1 \Myr$), which is approximately the life-time of the 
supermassive star and quasi-star phase. We postpone to a future work a detailed
 study of the observability of individual events of massive seed formation, but
this would be approximately the number density of supermassive stars/
quasi-stars from galaxy mergers as predicted by our
model.

\subsection{Black hole properties} \label{sec:properties}

We now look at the global properties of the supermassive black holes descendant  of
massive seeds, comparing them both with observations and with the
properties of black holes with a light seed progenitor.

 We start by looking at the evolution of the mass function, for black holes both with light and 
 massive seed progenitors (Figure
\ref{fig:MF_BH}). For the latter, we show the results obtained
assuming both a $1  \pc$ reservoir of gas and a $0.1 \pc$ reservoir. 
In the first case (upper panel), the black holes from massive seeds (red curves)
dominate almost the entire mass function at $z \sim 6$ and above; at $z=4$ the
mass function is already dominated by light seeds descendants (blue curves), except for the most
massive end. For a smaller reservoir cloud ($0.1  \pc$), the direct collapse
black holes can grow to larger masses after their formation: in this case, a substantial
number of black holes with masses above $10^8 \Msun$ is already present at $z=6$, and
the direct collapse  black holes dominate the massive end of the mass function to $z
\sim 2$. 
In the local universe, in a wide mass-range, the number density of black holes
from light
seeds is a couple of order of magnitude higher than the number density of black
holes formed from direct collapse. Only at the very massive end (above
$10^{9} \Msun$), direct collapse black holes are about $10\%$ of the total population
for the $1  \pc$ reservoir case. For the case in which the gas reservoir is
$0.1  \pc$, at $z=0$ and above $10^8 \Msun$ the massive seeds descendants are only about one
order of magnitude less numerous than the light seeds descendants with the same final mass.
Around $10^{10} \Msun$, the black hole with a massive progenitor are almost as numerous
as the light seeds descendants. 
In the same figure, at $z=0$, the 
the grey band indicates observational estimates from \citet{shankar04}, and, at $z=2$ and $z=4$
from \citet{merloni08}. Finding the model parameters that better match the black hole mass 
function is beyond the goal of this paper, given also the still large uncertainties in the estimates of the mass function at high redshifts.  We note, however, that the $z=0$ mass function is very 
well matched by the current model, as also reported by \citet{marulli08}. There is some 
discrepancy with the mass function estimated by \citet{merloni08} for $z=4$, but we see that a 
smaller  reservoir cloud could help increasing the mass function at the high-mass end, and 
possibly other channels of black hole growth, such as secular instabilities, could help increasing  the normalization of the predicted mass function.  
\citet{willott10} attempted to estimate the mass function of the black holes powering the very luminous
quasars observed at $z\sim6$.  When we compare with their results, we find a rough agreement at scales below $10^8 \Msun$. At higher masses, where their mass function is best constrained, we unfortunately do not have model predictions, as the volume of the Millennium Simulation is too small to sample the population of rare massive black holes at those redshifts. 

Finally, Figure \ref{fig:MF_BH} also shows the evolution of the active black
hole mass function, where we define a black hole as active if it is accreting
at a rate higher than $10 \%$ of its Eddington rate. 
 Essentially all black holes with a massive progenitor are actively growing at   $z=6$, but the fraction of active black holes strongly
drops with decreasing redshift (orange, dashed curves). This indicates that the
most massive black hole seeds 
  formed and grew at early time. Indeed the most massive end of
the direct collapse black hole mass function does not evolve much below $z=3$.
On the other end, the  descendants of light seeds are strongly growing at intermediate
and low redshifts, and, by $z=0$, they completely dominate the mass function also at
the highest masses. This is likely due to the fact that the most massive galaxies (and dark matter haloes), mainly assembled at late times \citep[][]{delucia07, angulo12}, when the probability of satisfying our constraints for the formation of a massive seed is lower. 
It is not obvious how the accretion rates of massive seeds accreting from the gas reservoir translate into luminosity, 
we thus do not attempt here to predict the quasar luminosity function from the active black hole mass function shown here.

\begin{figure}
\begin{center}
        \includegraphics[width=0.45\textwidth]{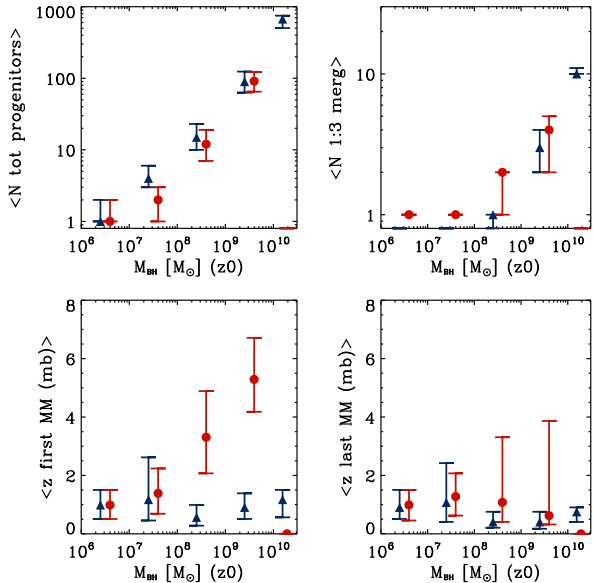}
        \caption{Merger histories of the galaxies hosting light seed descendants
(blue triangles) and massive seed black holes in the model with a $1 \pc$
reservoir (red circles),
as a function of black hole
mass at $z=0$. The top left panel shows the average number of progenitors
(mergers), while the average number of major mergers is shown in 
top right panel.The bottom panels show the average redshift of the first major
merger (bottom-left) and the average redshift of the last major merger
(bottom-right). The error bars bracket the $25$th and the $75$th percentile range.}
        \label{fig:NMergers}
\end{center}
\end{figure}

Thanks to the cosmological framework in which we work, we can directly compare the merger history of black holes with
light and massive seed progenitors.
Figure \ref{fig:NMergers} shows several properties of the merger history of
 black holes that are descendants of 
light seeds and direct collapse black holes that had a
first accretion phase assuming a
reservoir of $1 \pc$. 
All
quantities analyzed are plotted as a function of the black hole mass
at $z=0$; as both the descendants
of light seeds and massive seeds (with $r_r = 1\pc$) show a relatively tight relation with galaxy 
mass (in Figure \ref{fig:BH_Bulge_3z} we show the relation with bulge mass), studying the 
history of black holes at fixed mass, implies fixing the mass of the host galaxy.  
 The top panels indicate the average number of progenitors\footnote{We note that the absolute number of progenitors per galaxy 
 (or black hole) depends on 
 the resolution of the simulation, but the relative differences in the merger history of light and massive 
seed progenitors is insensitive to the resolution limits.} (or
mergers) that contributed to build up the black holes of final mass
indicated in the x-axis.
Across all masses, light
seed descendants
and direct collapse black holes seem to have had a similar total number of progenitors (top-left panel).
 The most massive black holes are sitting in galaxies that had hundreds of
  progenitors. Of all the mergers, only few of them are classified
as major (with mass ratio above $1:3$). While there is a small evidence that the
host galaxies of  black holes from massive seeds experience more major mergers than the host
galaxies of light seed descendants, the difference is not large, and the overall trend
as a function of mass is the same for the two populations (top-right panel).
The bottom panels show the average redshifts of the first and last major mergers.
While there is only a mild difference in the typical redshift of the last major merger, with galaxies hosting light seeds having the last merger at lower redshift than galaxies hosting  massive seeds   (right bottom panel), there is a clear difference in the
median redshift of the first major merger between the most massive light seed
descendant and
 massive seeds descendant of the same mass (left bottom panel):  black holes originated from direct
collapse are sitting in
galaxies that experienced a first major merger at much higher redshifts than
the galaxies hosting light seeds and, as we have seen in Fig.
\ref{fig:constraints_z_evol}, most major mergers at $z$ above $4$ potentially satisfy the
conditions we imposed for the formation of a direct collapse black hole seed.

\subsection{Black holes and their environment} \label{sec:environment}

Given that we are studying black hole seeds in a full galaxy-formation model applied to the outputs of the Millennium Simulation, we have the perfect framework to study not only the  cosmological evolution of black holes, but also the large-scale environment in which black holes grow. 
Once black holes have experienced exponential growth to become the ``mature'' objects that power AGN, it is impossible to recover information on the properties of their seeds, but the large scale environment might help us reconstruct the history of the hosted black hole.

\subsubsection{Scaling relations}

\begin{figure*}
\begin{center}
        \includegraphics[width=0.9\textwidth]{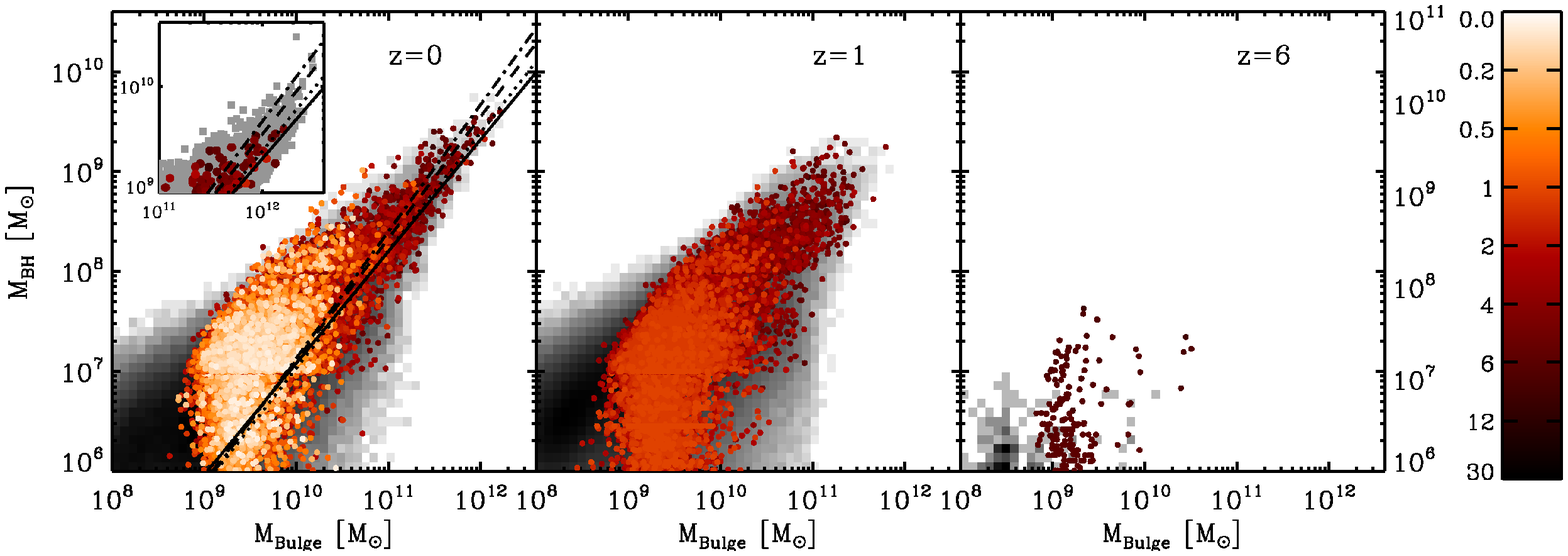}
        \includegraphics[width=0.9\textwidth]{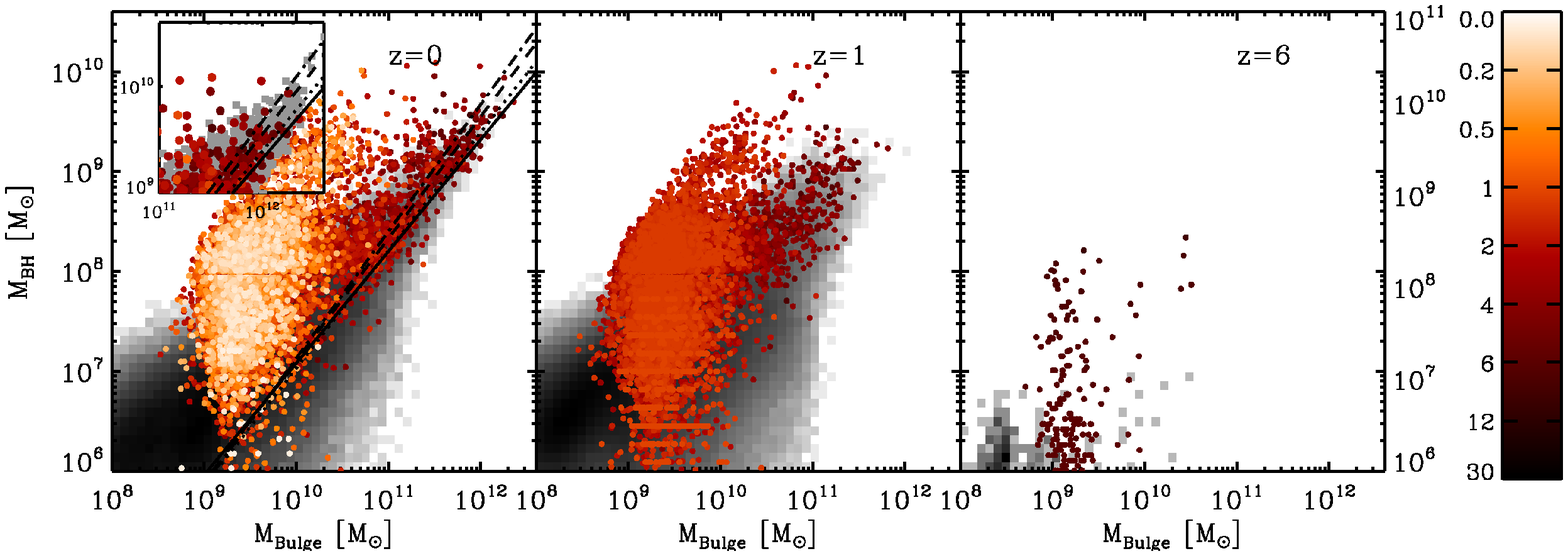}
        \caption{Relation between the black hole mass and the bulge mass at
three different redshifts, as indicated on the panels. The
grey area indicate the relation for light seeds descendants. The red points
indicate the position on the relation for the black holes descendant of  massive seeds. Darker red corresponds to higher formation redshift, as indicated in the
side color bar. In the top panel the massive seeds formed assuming a radius  of
$1 \rm{pc}$ for
the nuclear gas reservoir, whereas in the bottom panel the reservoir has been assumed to
have a radius of $0.1 \rm{pc}$. The solid lines at $z=0$ show some fit from
observational data \citep{haring04, gultekin09, hu09}.  }
        \label{fig:BH_Bulge_3z} 
\end{center}
\end{figure*}

Figure \ref{fig:BH_Bulge_3z} shows the relation between black hole 
and bulge mass at three different redshifts. The white-black area shows
the position of the light seeds descendants  on the $\MBH - \MBulge$ plane, where
darker pixels indicate a higher number density of objects. The 
success of the prescription of black hole growth described in Eq. \ref{eqn:quasar_merg_z} to 
reproduce the observed $\MBH-\MBulge$, especially at high masses, has already been discussed 
in previous works 
 \citep{croton06a, marulli08} (here the scatter is larger than previously shown as we include in the 
 figure also satellite galaxies). For the growth of massive seeds soon after formation, we do not 
 impose any preferred scaling with the mass of the host galaxy or dark matter halo, but rather a 
 very simple self-regulation mechanism,  where the main unconstrained
  parameter is the size of the reservoir cloud from which the new massive seeds can grow (see
  Eq.  \ref{eqn:DM_DC}). In Figure  \ref{fig:BH_Bulge_3z} the red dots indicate where massive 
  seeds descendants sit in the relation,  for a $1 \pc$ reservoir (upper panels),
and $0.1 \pc$ reservoir (lower panels), with the symbols color-coded depending on the 
 redshift of formation of the seed.  The oldest
 black holes from direct collapse formed around $z=12$,
and they are among the most massive black holes at $z=0$, but there is also a
large number of black holes that formed at later times and quickly accreted to the
highest masses. In the $1 \pc$ reservoir case, as soon as they are formed, the
    seeds sit on the relation, or slightly above it. It is interesting that the simple self-regulation mechanism for newly-formed massive seeds is able to bring them close to the relation, when $r_r=1\pc$ is assumed.
During
subsequent mergers, the black holes from massive seeds grow along the $\MBH - \MBulge$ following the prescription for 
light seeds (Eq.  \ref{eqn:quasar_merg_z}), 
and, by $z=0$, the most massive black holes show very little scatter. 
The picture is different when a $0.1 \pc$ reservoir cloud  is assumed: the seeds are allowed 
to grow to higher masses during the
first phase of accretion, and only the black holes that experience many subsequent
mergers head towards the relation. At $z=0$, the most recently formed massive
seeds are well above the relation. Assuming that the value of $r_r$ would be in
between the two values adopted here, our model predicts few outliers in the
relation; detecting such outliers, however, might be
observationally difficult, given the challenges in measuring directly 
black hole masses and given the low space-density of black holes descendant of
massive seeds.

\subsubsection{Properties of the host galaxies and haloes}
\label{sec:hosts}

\begin{figure}
\begin{center}
        \includegraphics[width=0.48\textwidth]{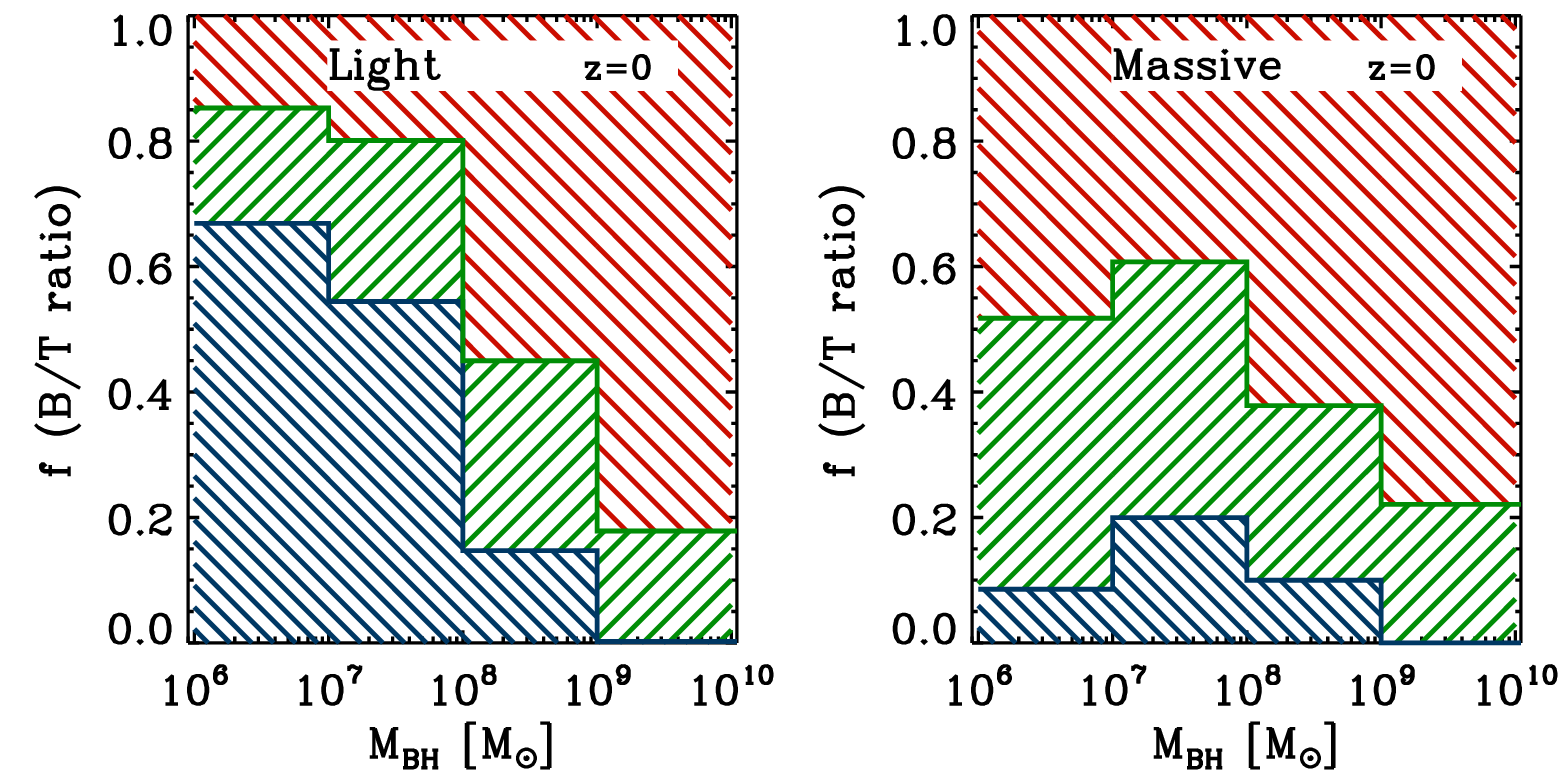}
        \includegraphics[width=0.48\textwidth]{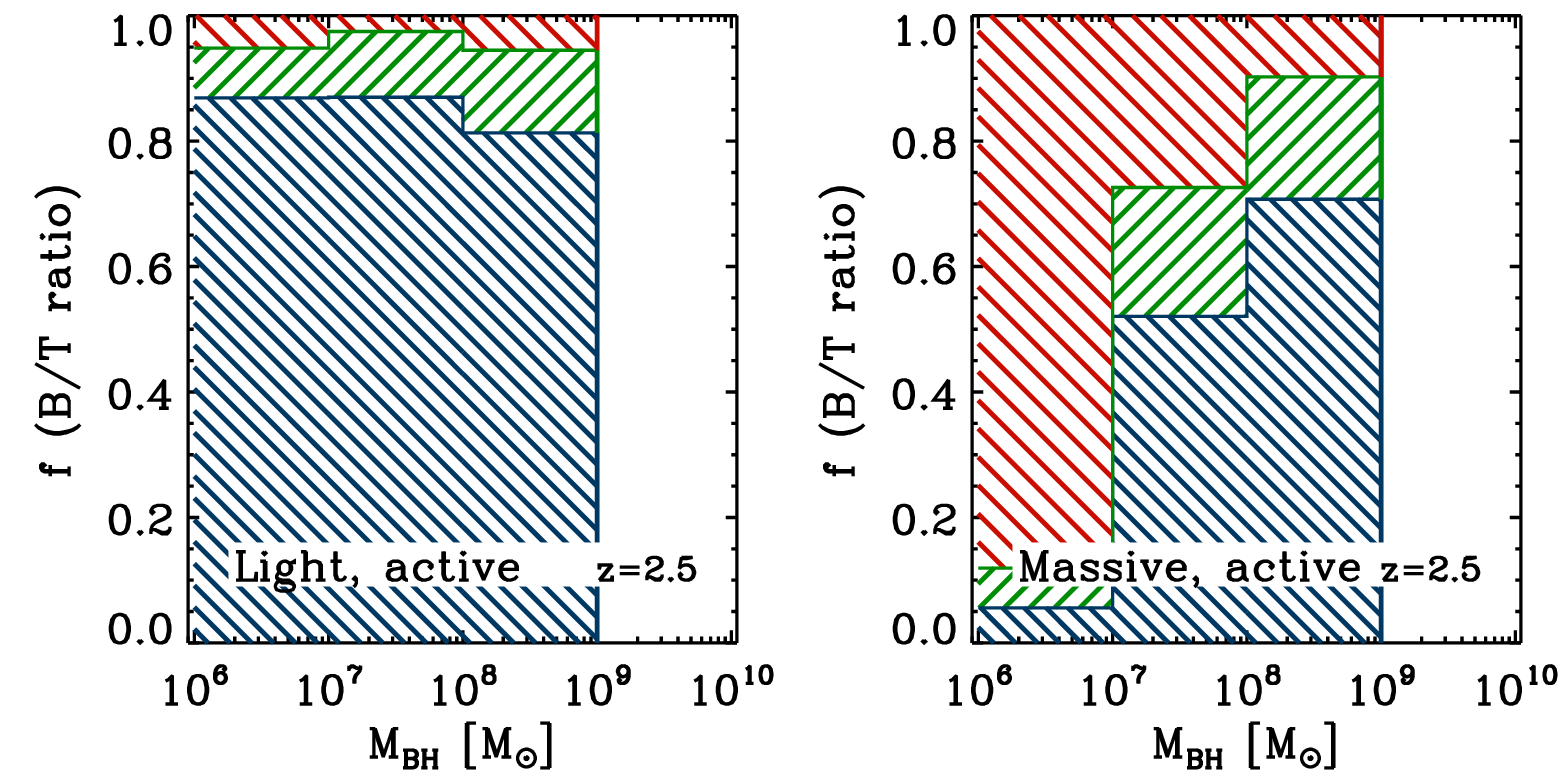}
        \caption{Morphological distribution of the galaxies hosting the
	descendants of light or massive 
        seeds (left and 
        right panels respectively), at $z=0$ (upper panels) and at $z=2.5$ (lower panels). At each 
        black hole mass, the area of different colors indicate the relative contribution  of galaxies with 
        different  morphologies, defined through the bulge-to-total ratio:  blue
refer to discs or extreme late-type  ($B/T <0.3$), green
 to normal spirals ($0.3 < B/T < 0.7$)
and red to elliptical galaxies ($B/T >0.7$). At $z=2.5$, only active black holes
 are considered. }
        \label{fig:galaxy_properties}
\end{center}
\end{figure}

In Figure \ref{fig:galaxy_properties} we show the morphological properties of the galaxies hosting
descendants of light and massive seeds.  The top panels show the distribution in
bulge-to-total ratio of the galaxies hosting black holes with light seed
 (top-left) and massive
seed  (top-right) progenitor as a function of the black hole mass at $z=0$.  
For the descendants of light seeds, small black holes live predominantly in disk-dominated galaxies (blue 
 area), while the majority of massive black holes sit in bulge dominated galaxies 
(red area). For massive seed descendants, the morphology distribution of host galaxies 
is more constant across black hole mass, with an essentially negligible amount of black holes in 
purely disk galaxies. In the lower panels the same morphology distributions of host 
galaxies is shown for objects at $z = 2.5$ and, in this case, only galaxies hosting an active black 
hole (accreting at a rate higher than $10\%$ of Eddington) are included in the calculation. At this redshift, very few black holes above 
$10^9 \Msun$ are 
active (see the active mass function, shown by the dashed curves in Figure \ref{fig:MF_BH}), so we exclude them from the present calculation. Black 
holes from 
light seeds sit mainly in disk-dominated galaxies, and only a negligible amount is in spheroids. 
The same holds for the active descendants of  massive seeds in the $10^7 - 10^8 \Msun$, but the smaller black holes (mainly recently-formed seeds) are  
essentially only hosted by ellipicals.
Our model clearly predicts quite different morphological properties for the galaxies hosting light and massive seeds remnants, in particular for black holes of intermediate-small mass scales.  Galaxy morphology could then contain important information on the origin of the hosted black holes.

\begin{figure}
\begin{center}
        \includegraphics[width=0.48\textwidth]{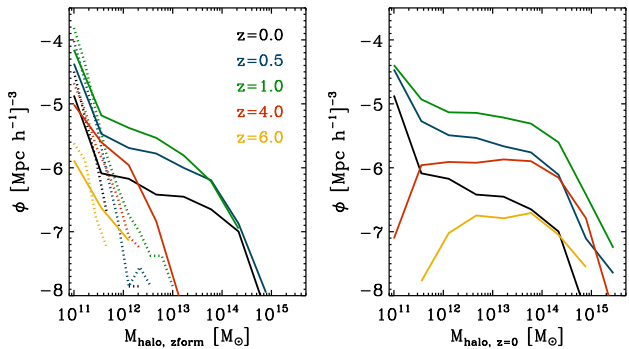}
        \caption{ {\bf  Left panel:} At various redshifts, mass function of the dark matter subhalos (dotted lines)
and  haloes (solid lines) 
hosting newly-formed massive seeds. {\bf Right panel:} local mass function of the haloes hosting the descendants of the massive seeds formed at the redshifts of the left panel. }
        \label{fig:MF_hosts}
\end{center}
\end{figure}

We now  look at the properties of the  dark matter  haloes hosting massive seeds. 
In the left panel of Figure \ref{fig:MF_hosts} we show, at various redshifts, the mass function of the   
haloes (solid  lines) and subhaloes (dotted lines)
hosting newly-formed massive seeds (for central galaxies we take the virial  mass of the halo as subhalo mass).  
The mass functions clearly peak at $10^{11} \Msun$, as 
expected, given that  we impose that value as the minimum  halo mass 
where massive seeds can form.  As in central galaxies, according to our definition, the subhalo and halo masses are equivalent, 
 the two mass functions trace each other for massive seeds forming in central galaxies. This 
seems to generally be the case for massive seeds forming in haloes smaller than  $\sim 10^{12} \Msun$. 
At larger 
masses, while the subhalo mass function drops quickly, the  halo mass function flattens out and, at 
lower redshifts, reaches very high masses: these are massive seeds forming in subhaloes that are  
the hosts of satellites galaxies in larger haloes, that reach the masses typical of clusters. We find, in fact, that at $z=0$ about $20\%$ of new-born massive 
seeds are in satellite galaxies.

In the right panel we show in which  haloes the descendants of these massive seeds are 
today: seeds formed at very high redshifts sit in very massive  haloes, with only a small 
fraction in haloes below $10^{12} \Msun$. The descendants of seeds forming at more recent times 
are also  sitting in massive clusters, but there is still a fraction of the population in smaller haloes.

\subsubsection{Clustering}

It is well known that the space distribution of dark matter haloes and galaxies
can be different from the
distribution of the underlying dark matter  \citep{kaiser84, bardeen86}. 
The two-point auto-correlation function\footnote{The {\em two-point spatial autocorrelation
  function} $\xi(r)$ for a given class of objects is defined as  
 the excess probability for finding a pair  at a
distance {\em r}, in the volume elements ${\rm d}V_{1}$ and ${\rm
  d}V_{2}$:
${\rm d}P=n^{2} \left [ 1+ \xi(r) \right ] {\rm d}V_{1} {\rm d}V_{2}$,
where $n$ is the average number density of the set of objects under
consideration  \citep[e.g.,][]{peebles80} } provides information on how strongly
 a given class of objects is clustered at a given scale, and from the
amplitude and shape of this function is possible to extract important
information on
the environment of the objects analyzed. 

One of the main advantages of studying galaxy evolution in models run on dark matter simulations rather than extended Press-Schechter models, is the possibility of studying the clustering of  a targeted set of objects, using the distribution 
of the dark matter haloes in which the objects reside.

At any given time, the most massive dark matter haloes are the most rare and
biased\footnote{At any given scale, the bias parameter indicates how more (or less) clustered
objects are with respect to the dark matter.}
objects, corresponding to the highest peak of the dark matter density field. 
While the clustering amplitude of dark matter haloes
depends mainly on halo mass and only weakly on other properties, such as
assembly history, concentration, recent mergers \citep[e.g., ][]{gao05, wechsler06, angulo08, bonoli10}, the
clustering of galaxies depends strongly not only on mass, but also on 
galaxy properties such as color and surface density:  for example, at fixed stellar mass, red and passive galaxies
cluster more strongly than blue, star forming galaxies \citep[e.g.,][]{li06b}, which is a consequence of galaxies of a given stellar mass, populating different 
dark matter haloes.

\begin{figure*}
\begin{center}
        \includegraphics[width=0.8\textwidth]{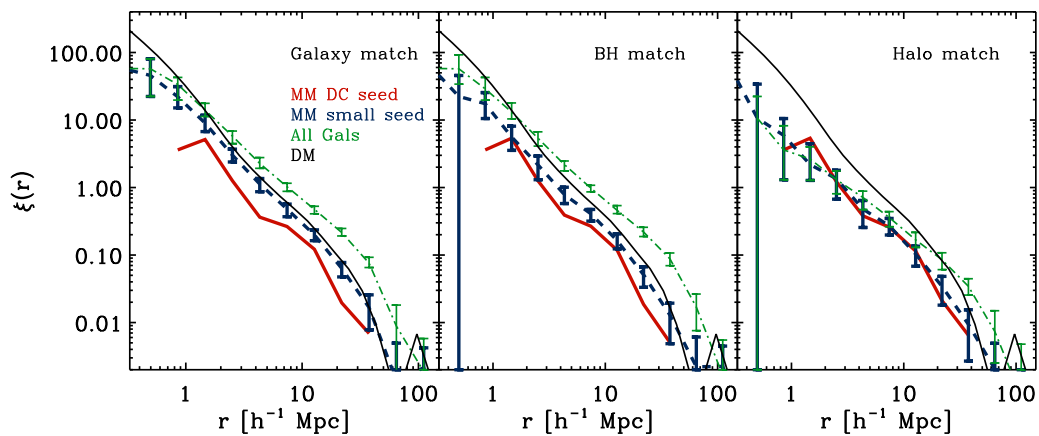}
        \caption{Two-point auto-correlation function of galaxies hosting the
formation of massive black hole seeds from major mergers (red curves) at $z=0$. The blue dashed curves and the green
dotted-dashed curves show the correlation function of, respectively, other major
mergers and the whole galaxy population with matching stellar mass (left panels),
black hole mass (central panels) and dark halo mass (right panels). The solid
line indicated the autocorrelation of the dark matter in the Millennium simulation. }
        \label{fig:clustering}
\end{center}
\end{figure*}

Here we want to study the clustering properties of galaxies that host the
recent formation of a massive black hole seed, to further gain insights
on the large-scale environment of these events. 
In Figure \ref{fig:clustering} we show the $z=0$ two-point correlation
function of galaxies that experienced a major merger that led
to the formation of a massive seed (red solid curves). For comparison, we also show the
two-point correlation function of all other major mergers at the same epoch that
did not satisfy the conditions for direct collapse (blue dashed curves),
randomly extracting, from the entire population, a subsample that matches the
massive-seed population in number of objects and in the distribution of stellar
mass (left panel), black hole mass (central panel) or halo mass (right
panel). Using the same matching criteria, we also randomly extract subsamples
from the entire galaxy population, and the correlation function of these objects
is indicated by the green dotted-dashed lines. To calculate the
uncertainty in the auto-correlations due to random sampling, for each matching
criteria, we selected $100$ random subsamples from the parent populations, and
the error bars bracket the $10$ and $90$ percentiles of the distribution. 
We find that galaxies that
recently experienced a direct collapse event are significantly less-clustered
than the rest of the major-merger population and the entire galaxy population,
when these are matched by stellar mass or black hole mass (left and central panels). 
Moreover, all recent
mergers (both the ones that lead to a massive seed and the ones that preserve
the small seed) are anti-biased with respect to the dark matter
distribution. On the contrary, the random samples extracted from the entire galaxy population is slightly biased.  
Clearly, this indicates that galaxies with similar stellar mass
or black hole mass cluster differently depending if they experienced a recent major
merger. This is in line with the observational results indicating that blue-active galaxies
are less clustered than red-passive ones. When matching samples by halo mass,
these differences essentially vanish. We find, in fact, that, at fixed stellar
mass, galaxies with a recently formed massive seed tend to be found in less
massive haloes than the average galaxy and, when we match out samples by
halo mass, the differences in clustering get erased as
halo clustering depends primarily on mass. 
The smaller clustering amplitude of recent mergers (and, to an even larger
extent, of mergers that lead to direct collapse), indicates that, in the local
universe,  anti-biased galaxies are the possible sites for
finding on-going events of direct collapse. We find that these are mainly galaxies that lived in
isolation for most of their lives, and that only recently experience encounters
with other objects with similar properties.

We also looked at the clustering behavior of galaxies hosting  newly-formed
massive sees at 
high redshift, but we could not find a signal as strong as the one showed above:  all samples show 
a similar clustering
amplitude, independently on the assumed matching property.
Unlike the $z=0$ results, we also find that at high 
redshifts galaxies hosting a new massive seed show a 
much higher correlation function than the underlining dark matter 
distribution, which is expected, given that haloes above  $10^{11} \Msun$ (which is the imposed minimum mass in which massive seeds can form) reside in higher and higher density peaks as redshift increases. This is in line with quasar measurements at high
redshifts ($z > 2$), which indicate that bright quasars are a highly-biased
population, living in high density peak \citep[e.g.,][]{shen07,shen09}.  In our model the black holes descendants of massive seeds  dominate the
massive-end of the mass-function at very high redshift (see Figure \ref{fig:MF_BH}),
so these could be the objects powering the brightest highly-clustered quasars.

\section{Summary and Conclusions} \label{sec:conclusions}

In this paper we have introduced a new scenario for the formation of massive black
hole seeds. Based on the results of the set of
hydro-simulations from 
\citet{mayer10}, we tightly link the formation of a massive black hole to the
major mergers of gas-rich disk-dominated massive
galaxies with no pre-existing massive black hole at their center. Such mergers
can, in fact, easily channel a lot of gas at the center of the merger remnants,
forming a sub-parsec scale cloud which will likely lead to a massive black hole
seed either through direct gravitational collapse or, more likely, through a supermassive star/
 quasi-star phase. In mergers of this kind, in fact, the accretion rate to the center are seen to be 
 order of magnitudes higher than in isolated protogalaxies.
 
  We developed a formalism to track such events in galaxy
formation models to predict the evolution and environment of massive seeds across cosmic time. 
Rather than tuning the model free parameters that control the formation of massive seeds with observational data, we set their value using the findings of the hydro-simulations of M10, so that all our results on massive seeds can be considered genuine predictions of theoretical models.   
These are our main results:
\begin{itemize}
\item{At redshifts above $z\sim 3-4$ almost all major mergers of galaxies
residing in haloes of at least
$10^{11} \Msun$ meet the imposed conditions for the formation of a massive black
hole seed. At lower redshifts the fraction of major mergers able to produce a
massive seed strongly drops, and, by $z=0$, the fraction has reduced to $\sim
20\%$. This dropping is due to the sharp decrease in the probability of having a
major
merger which involves two disk-galaxies with a still small black holes.  
}
\item{Massive black hole seeds can dominate the massive end of the mass function
above $z\sim2$ or $z\sim 4$ depending on the radius of the gas reservoir from
which they accrete just after their formation. Newly-formed massive seeds are, in fact, allowed to accrete from  
the surrounding gas until their feedback energy is able to unbind it.  As the binding energy of the gas
reservoir depends on its physical size (at fixed mass), massive seeds grow to larger masses  in the model runs where 
the reservoirs are assumed to be smaller. }
\item{Massive black hole seeds soon after formation sit on or above the
$\MBH-\MBulge$ relation, depending mainly on the size of the reservoir assumed. We
generally expect a fraction of descendants of massive seeds to still be above the
relation in the local Universe, but this population might not be easy to detect,
given the predicted small number density of these objects and given the
difficulties in measuring black hole masses directly.}
\item{Massive seeds evolve very rapidly at high redshift, but do not grow significantly in the local universe.
 Of the most massive black holes today, the ones descending from a massive seed  
 sit, in fact, in galaxies that had a first major merger
very early. On the contrary, the massive descendant of light seeds are 
  in  galaxies that had a first major merger relatively recently,
but grew very quickly to high masses. }
\item{While the most massive black holes today sit in bulge-dominated galaxies, independently if they descend from a light or massive seed, the morphology of the host galaxies of smaller black holes is very different for the descendant of light and massive seeds: black holes between $10^6 \Msun$ and $10^8 \Msun$ with a light seed progenitor  preferentially sit in disk-dominated galaxies while the descendants of massive seed in the same mass range are hosted by bulge-dominated galaxies. }
\item{Galaxies that host massive
seeds formed at very recent times have significantly lower clustering amplitude than  a random subsample of the 
global galaxy population with the same stellar mass distribution. While this can
be explained by the fact that, at fixed stellar mass, recent mergers take place
in smaller haloes, the observational signature of this effect should be very clear. }
\end{itemize}
Massive black hole seeds are a very attractive scenario alternative or, more likely, 
complementary, to models of light seeds from PopIII stars or the
collapse of nuclear star clusters. While massive seeds from metal-free
protogalaxies can
only form at very high redshift, in this work we have shown that massive seeds from galaxy mergers, which do not require metal-free gas but rely on a
different set of conditions, can  form both at high redshift and 
at more recent times.  While we have considered a purely hydrodynamical formation scenario for the massive seeds, which, at small scales, would
be compatible with the quasi-star model of Begelman and collaborators, the results of our model should hold even if the ultimate collapse into
a seed would take place in a different way, such as involving the core collapse of a massive nuclear cluster
of normal stars \citep{davies11}, as long as the formation of the  
precursor supermassive gas cloud occurs  under the conditions that we have considered here.
Whether the ``birth'' of such black holes could be observed directly, is still very uncertain. A promising detection channel has recently been suggested by  \citet{czerny12}, who claim that quasi-stars might emit 
jets whose gamma-ray emission might account for the unidentified gamma-ray sources.  The quasi-star phase, which has an emitted
spectrum very similar to that of a red giant star, might be detectable with JWST at high redshift given
the relative high frequency of seed formation events expected in models that rely on the merger rate
of galaxies such as in our scenario or in other recent direct collapse models
\citep{volonteri10}. 
On the other end, seed formation events happening at low redshift might be eventually identified by exploiting the information presented in this paper
on the environment and nature of their host galaxies. Indeed, if they take place they should occur in underdense regions, perhaps
in voids or at least field-like environments, where gas-rich, massive disk dominated spirals are common 
(good examples of such galaxies are indeed those in the local $8 \Mpc$ Volume
\citep[e.g.,][]{kormendy10}.
 We will dedicate a follow-up paper to a detailed analysis of the possible detectability of individual events at both high and low redshift.

\section*{Acknowledgments}
We thank Marta Volonteri, Mitch Begelmann, Takamitsu Tanaka,  Raul Angulo and Simon White for useful discussions and
comments on the manuscript.
SB particularly thanks Simon White also for the generous hospitality at the 
Max Plank Institute for Astrophysics, where a large fraction of this work has been carried on.

\bibliographystyle{mn2e}
\bibliography{SAM_seeds}

\label{lastpage}

\end{document}